\documentclass[12pt]{iopart}

\usepackage{graphicx}

\begin{document}

\newcommand{\eu}{$^{153}$Eu$^{3+}$:Y$_2$SiO$_5$}
\newcommand{\tranz}{$\pm|3/2\rangle_g \rightarrow \pm|3/2\rangle_e$ }
\newcommand{\trann}{$\pm|5/2\rangle_g \rightarrow \pm|3/2\rangle_e$ }
\newcommand{\trans}{$\pm|1/2\rangle_g \rightarrow \pm|5/2\rangle_e$ }
 \newcommand{\transition}{$^7$F$_0 \rightarrow ^5$D$_0$-transition }

\title[Atomic frequency comb memory with spin wave storage in \eu]{Atomic frequency comb memory with spin wave storage in \eu}

\author{N Timoney, B Lauritzen, I Usmani, M Afzelius and N Gisin}

\address{Group of applied physics, University of Geneva, CH-1211 Geneva 4, Switzerland}

\ead{mikael.afzelius@unige.ch}
\begin{abstract}
\eu $~$is a very attractive candidate for a long lived, multimode quantum memory due to the long spin coherence time ($\sim$15 ms), the relatively large hyperfine splitting (100 MHz) and the narrow optical homogeneous linewidth ($\sim$100 Hz). Here we show an atomic frequency comb memory with spin wave storage in a promising material \eu $~$, reaching storage times slightly beyond 10 $\mu$s. We analyze the efficiency of the storage process and discuss ways of improving it. We also measure the inhomogeneous spin linewidth of \eu$~$, which we find to be $69 \pm 3$ kHz. These results represent a further step towards realising a long lived multi mode solid state quantum memory.
\end{abstract}
\submitto{\jpb}

\maketitle
\section{Introduction}

Quantum communication \cite{Gisin2007a} provides resources and capabilities, such as quantum key distribution \cite{Gisin2002}, that are not possible to obtain using classical communication. A major challenge to quantum communication, however, is to overcome the inherent losses of quantum channels, e.g. optical fibers. A solution to the problem of long-distance quantum communication is the quantum repeater \cite{PhysRevLett.81.5932,Duan2001,Simon2007,RevModPhys.83.33}, which in principle can work over arbitrary distances.
To implement a quantum repeater, quantum memories, effective delay lines of variable duration, are required. Atom based memories are attractive candidates, indeed much research has been done in recent years to implement them \cite{Simon2010,LPOR:LPOR200810056}. In addition to faithfully reproducing the input mode and storing for potentially long times \cite{PhysRevLett.95.063601, Zhao2009}, memories should also have a multimode capacity \cite{Usmani2010, 1367-2630-13-1-013013} and an on demand readout \cite{PhysRevLett.104.040503} to realize a good quantum memory for quantum repeaters. The quantum memory based on an atomic frequency comb (AFC) \cite{PhysRevA.79.052329} has the potential of achieving these ambitious goals in one memory.

An AFC memory is one which is based on an inhomogeneously broadened medium which has been tailored to contain a frequency comb of narrow peaks (teeth) of atomic population. The frequency spacing between the narrow peaks ($\Delta$) dictates the storage time of the memory: coherent re-emission of the input mode, an echo, is seen in the same forward direction a time 1$/\Delta$ after the input mode. Such an echo is referred to as a two level echo in the remainder of this paper. The efficiency of the echo in the forward direction, assuming a comb which can be described by a sum of Gaussian functions, is given by \cite{PhysRevA.79.052329}

\begin{equation}
\eta \approx \left(\frac{d}{F}\right)^2e^{-7/F^2}e^{-d/F}e^{-d_0}
 \label{eq:GaussEff}
 \end{equation}

 \noindent where d ($= \alpha L$, $\alpha$ is the absorption coefficient and $L$ the sample length) is the optical depth and F is the comb finesse. More specifically F = $\Delta/\gamma$, where $\Delta$ is as before and $\gamma$ is the comb tooth full width at half maximum (FWHM). Here we also include an additional loss factor (the last factor) due to an absorbing background $d_0$, which often occur due to imperfect preparation of the comb. Note that although we consider Gaussian shaped teeth here, which fit well with our experimental data, other shapes have been considered elsewhere \cite{Chaneliere2010,PhysRevA.81.033803}. Important results have been obtained using such a two level AFC scheme, where a heralded single photon was stored in a crystal \cite{Clausen2011, Saglamyurek2011} or heralded entanglement was generated between two crystals \cite{2011arXiv1109.0440U}. Two level echo efficiencies using an AFC scheme of 15-25 $\%$ are seen in many experiments \cite{PhysRevLett.104.040503,PhysRevA.81.033803,Clausen2011,Sabooni2010,Amari2010}. In these papers and the work presented here, the inhomogeneously broadened medium is an inorganic crystal weakly doped with rare-earth ions. Such crystals placed in commercially available cryostats cooled to less than 4 K have impressive coherence properties on optical and spin transitions \cite{PhysRevLett.95.063601,PhysRevLett.72.2179}. We refer to \cite{LPOR:LPOR200810056} for a comprehensive review of these materials in terms of quantum memories. In this particular work we use \eu $~$, which we believe has the potential of fulfilling the requirements of a quantum memory for quantum repeaters \cite{Simon2007,PhysRevA.79.052329}.

\begin{figure}
\centering
\includegraphics[scale=0.35 ,angle = -90, trim = 220 150 130 150]{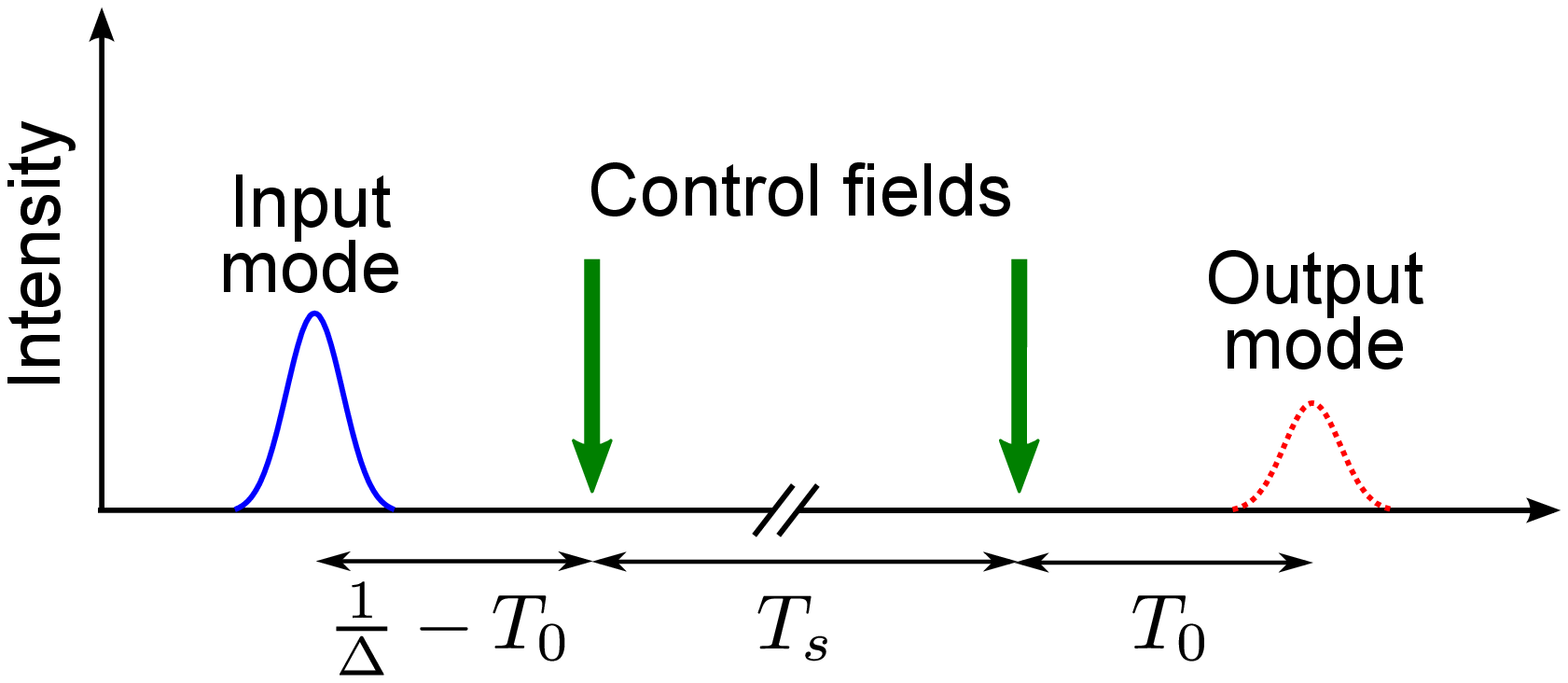} \\
\includegraphics[scale=0.2]{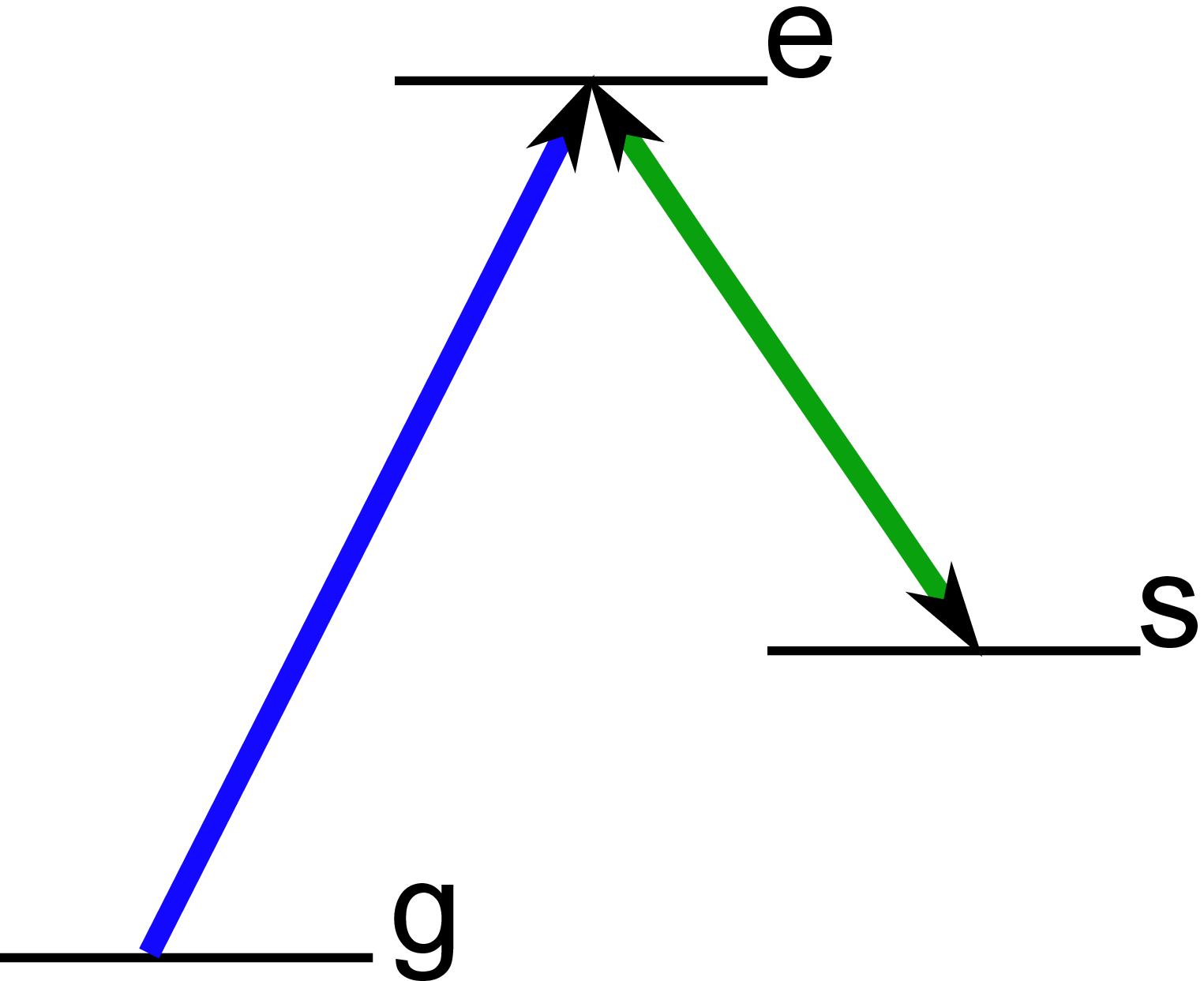}
\caption{An illustration indicating the time order of an AFC involving spin wave storage. The input mode is in resonance with the $|g\rangle \rightarrow |e\rangle$ transition, the control pulses are applied on the $|e\rangle \rightarrow |s\rangle$ transition. The time $1/\Delta$ is defined by the periodic frequency separation $\Delta$ of the teeth of the atomic frequency comb.}
\label{fig:AFCorder}
\end{figure}

The scheme described above, however, is not the complete AFC scheme as proposed in \cite{PhysRevA.79.052329}. The conversion of the optical excitation to a spin excitation is missing. This requires the presence of another ground state. For a complete scheme the input mode is followed by a control pulse which transfers the optical coherence between $|g\rangle$ and $|e\rangle$ to a spin coherence between $|g\rangle$ and $|s\rangle$. The time line and an illustration of the relevant levels is shown figure \ref{fig:AFCorder}. This control pulse 'stops the clock' of the predefined memory time of $1/\Delta$. A second control pulse 'restarts the clock' by reverting to the optical coherence between $|g\rangle$ and $|e\rangle$. The time between the control pulses is not predefined, such that the application of the second control pulse allows on demand readout of the memory. The explanation above is simplified since it does not take into account spin dephasing due to inhomogeneous broadening, which leads to a decay in storage efficiency as a function of spin storage time. This dephasing can, however, be compensated for by using spin echo techniques, allowing long storage times only limited by the spin coherence time.

An interesting aspect of the AFC scheme is that it is possible to implement a multimode memory in the time domain where the multimode capacity is not dependent on increasing optical depth \cite{PhysRevA.79.052329,Nunn2008}. Rather the limit on the number of temporal modes which can be stored depends only on the number of peaks in the atomic frequency comb. This is restricted, in turn, by the width of the AFC, generally limited by the hyperfine transition spacing and the smallest comb tooth width which can be obtained. Another restriction, of a technical nature, is imposed by the frequency bandwidth of the control pulses, which must spectrally cover the entire AFC spectrum. To transfer a large bandwidth, the control pulses would most likely take the form of chirped shaped pulses as discussed in reference \cite{PhysRevA.82.042309}.

To implement a full AFC scheme, the atomic element must have three levels in the ground state, so that the atoms not required for the comb can be spectrally separated from the memory in an auxiliary state. In terms of rare earth doped crystals, praseodymium and europium are interesting candidates. Their nuclear spin of I = $5/2$, results in three hyperfine levels at zero applied magnetic field \cite{PhysRevLett.95.063601}. A full AFC scheme was demonstrated for the first time in praseodymium \cite{PhysRevLett.104.040503}. The nature of the material used in these results places a limit of the number of modes to tens of modes. This limit is due to the hyperfine splitting of the material (approximately 10 MHz) and the optical homogeneous width of the material - 1 kHz.  Europium on the other hand has a larger hyperfine splitting (approximately 100 MHz) and an optical homogeneous linewidth of the order of 100 Hz, \cite{PhysRevLett.72.2179}, with such a material it should be possible to create a multi mode memory with at least an order of magnitude more modes \cite{PhysRevA.79.052329}. Also, larger frequency separation of the input and control frequencies will be useful for spectral filtering, a likely requirement for future single photon storage. The drawback of europium doped materials is the low oscillator strength, which results in low Rabi frequency of the control fields and low optical depth. Both these factors are serious limitations in our present experiment, as we will describe. We note, however, that a material with a small optical depth does not necessarily have a poor efficiency, as equation \ref{eq:GaussEff} would imply.  By placing the crystal in an impedance matched cavity it should be possible to achieve high efficiencies despite a low optical depth \cite{PhysRevA.82.022310, PhysRevA.82.022311}.

\section{Europium}

\begin{figure}
\centering
{\includegraphics[scale=0.5]{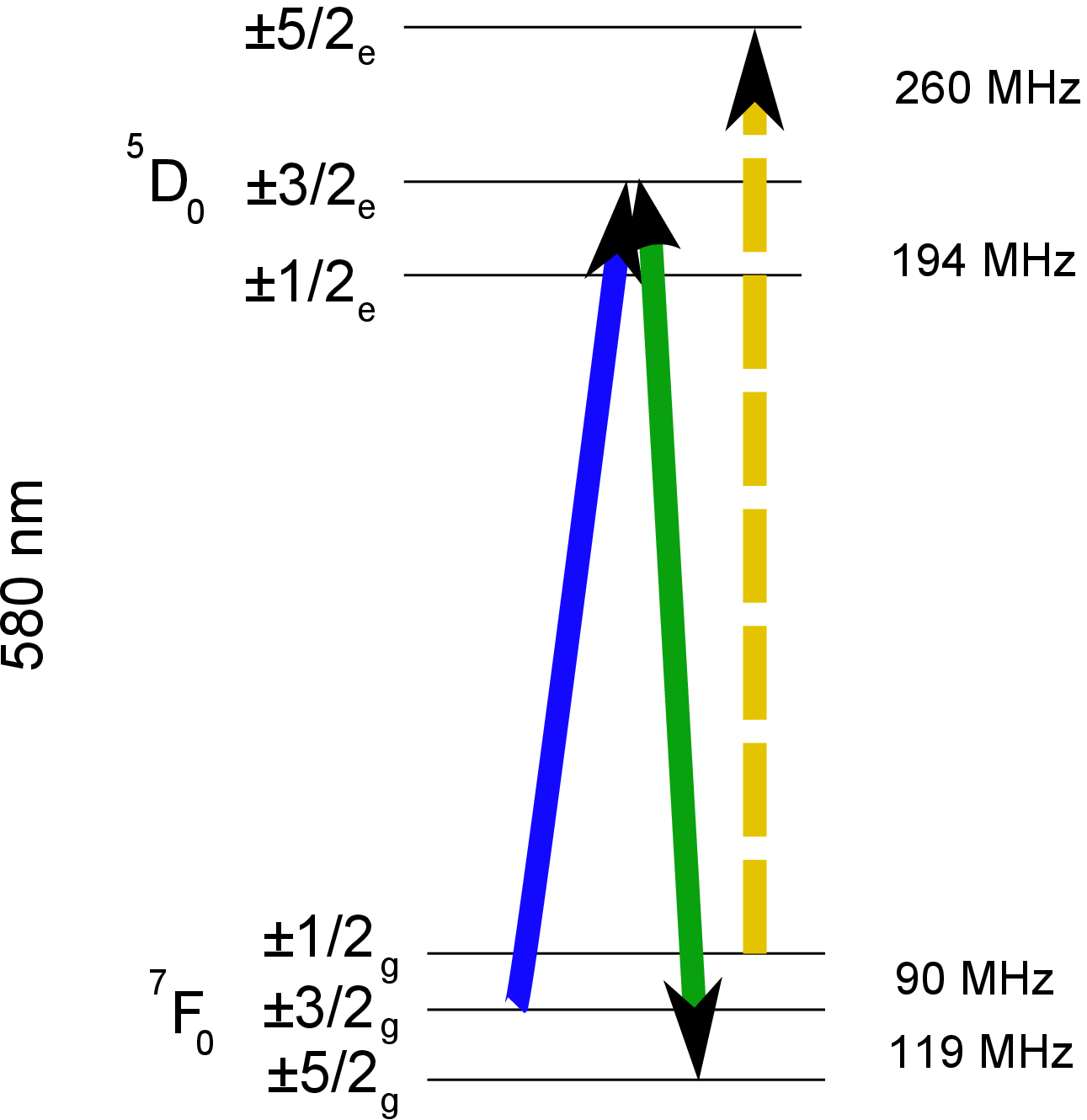}}
\caption{Hyperfine structure of the \transition in \eu. The blue and green arrows indicate the $\Lambda$-system used for this work.  Where the \tranz acts as the $|g\rangle \rightarrow |e\rangle$ transition for the input mode, and \trann the $|e\rangle \rightarrow |s\rangle$ transition for the control fields. Although a $\Lambda$-system containing two ground states is all that is required for the complete AFC scheme, a third ground state (an auxiliary state) is required to store the atoms which are not part of the AFC. For the $\Lambda$-system used in this paper the auxiliary state is $\pm|1/2\rangle_g$. The \trans transition is used in the comb preparation stage, as explained in the text.}
\label{fig:lambda_syst}
\end{figure}

Europium doped Y$_2$SiO$_5$ has been the subject of several spectroscopic studies, measuring for instance absorption coefficients, inhomogeneous and homogeneous broadenings and hyperfine level spacings \cite{PhysRevLett.72.2179,PhysRevB.68.085109,Yano1991,Yano1992}. Until recently, however, the ordering of the hyperfine levels and the transition branching ratios between these were unknown in ~\eu. We thus conducted spectroscopic investigations of an isotopically enriched \eu$~$ crystal \cite{submittedprb} to determine a suitable $\Lambda-$system in which to perform a complete AFC scheme. In this work we use one of the potential $\Lambda-$systems identified in our previous work, shown in figure \ref{fig:lambda_syst}. We use the stronger \tranz transition, with a greater optical depth, for the input mode, while the control fields are applied on the weaker \trann transition. It should be noted that from \cite{submittedprb} three other potential $\Lambda-$systems could be identified, which could all work as well or even better than the selected one (i.e. have larger oscillator strengths). This particular configuration was chosen for a first proof-of-principle demonstration due to available frequency shifts by our frequency shifters. Other configurations will be tested in future work.

Our crystal has a $^{153}$Eu$^{3+}$ doping level of 100 ppm, an inhomogeneous broadening of 700 MHz and an absorption coefficient of 1.2 cm$^{-1}$ on the \transition. We note that larger optical depths are possible with increased doping \cite{PhysRevB.68.085109}.

\section{Experimental description}

\begin{figure}
\centering
\includegraphics[scale=0.5]{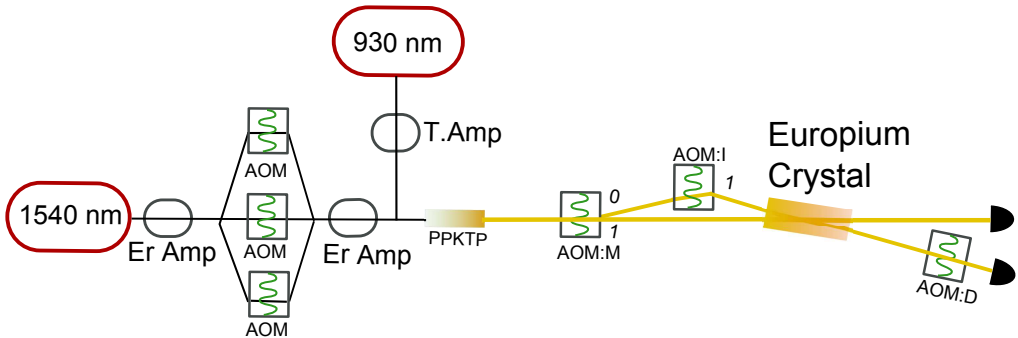}
\caption{This figure represents a very basic illustration of the experiment. The sum frequency process in the PPKTP waveguide generates the light at 580 nm from the 1540 nm and 930 nm diode lasers. The light from both lasers is amplified, a tapered amplifier is used for the light at 930 nm, and two erbium amplifiers for the light at \mbox{1540 nm}. The AOMs at 1540 nm select the atomic transition used. Two AOMs (AOM:M and AOM:I) at 580 nm before the europium crystal determine the amplitude and the duration of light in either the control or the input mode. These two modes overlap in the europium crystal. Further explanation of the experiment can be found in the text. }
\label{fig:schem}
\end{figure}

The \transition requires a light source at 580 nm. This wavelength is not covered by diode lasers at present, so we have chosen to generate it using sum frequency generation (SFG) of two wavelengths: 1540 nm and 930 nm. The non linear medium used is a PPKTP waveguide, which produces 110 mW of light at 580 nm at the output where there is 1.5 W of light at 1540 nm and 400 mW of light at 930 nm at the input. To reduce the losses of the light produced at 580 nm, the three frequencies required for the measurements shown in this paper are produced using accousto optical modulators (AOM's) on the 1540 nm light. This light is recombined in a 3x3 coupler, the polarization of each AOM can be controlled before it is combined, thus adjusting the polarization required for the waveguide.  The polarization control is suppressed in figure \ref{fig:schem}. The losses due to these AOM's are compensated with the amplifier after the 3x3 coupler, such recuperation could not be obtained at 580 nm.

Once the light is produced at 580 nm it passes through an AOM (AOM:M) which determines the intensity and duration of the light in the crystal. AOM:M is in double pass, which allows small frequency scans of at least 15 MHz to be made uniformly in intensity. In fact more than one spatial mode is required for the spin wave storage measurements performed in this paper. This is due to the low amplitude of the observed echo signal, compared to the strong control pulse. In addition, we observed that the control pulse causes free induction decay in its spatial mode, presumably due to off-resonant excitation, which will act as a background noise on the weak echo. To avoid this noise we use two spatial modes, one for the comb preparation and the control modes, another for the input pulse. In order to have fast independent control over each of the spatial modes we use two AOMs, one for each mode. The zero order from the first pass of AOM:M is thus diverted to another AOM (AOM:I) which operates at the same frequency. AOM:I is also in a double pass configuration. The resulting light from AOM:I and AOM:M cross in the europium crystal. Before the cryostat there can be up to 53 mW of light from AOM:M, depending on the radio frequency amplitude applied to AOM:M. This light is focussed in the crystal with a beam waist of 60 $\mu$m. The light from AOM:I is significantly weaker, it can only reach a maximum power of 3 mW. The light hitting the detector can be attenuated using AOM:D in single pass after the cryostat.

The light at 580 nm is frequency stabilized using a second continuous wave SFG source which is not drawn in figure \ref{fig:schem}. Light from both diode lasers is taken before the amplifiers, combined on a dichroic mirror and sent through a second PPKTP waveguide. A cavity with a free spectral range of 1 GHz and finesse of 600 is used in a Pound Drever Hall configuration. The cavity mirrors are separated by an invar spacer and the structure is temperature stabilized. The correction for the error signal is applied to \mbox{930 nm} laser. The frequency stabilization removed slow frequency drifts and narrowed the laser linewidth to slightly less than 50 kHz as measured by spectral hole burning.

The europium crystal itself is housed in a pulse tube cooler, where the cold finger has a temperature of 2.8 K. We observe spectral broadening of the teeth in the comb when the cryostat cooler is switched on compared to when it is switched off. The spectral broadening was very shot-to-shot dependent, indicating that it depends on where the measurement is performed within the pulse tube cycle (with a period of 700 ms). It is likely that this spectral broadening is caused by crystal movement in the same direction as the beam propagation, induced by the vibrations from the cryostat compressor or rotary valve. Although the exact physical mechanism leading to this broadening remains unclear, we believe it to be due to phase noise in the atom-light interaction induced by the modulation of the laser-crystal distance, leading to an effective laser line broadening. By triggering the experimental sequence on the vibrations in the cooling tubes using a piezo, the observed jitter on the width of a comb tooth is reduced by a factor of three, from 150 kHz to 50 kHz. We conclude this section by noting that the technical spectral broadening due to laser linewidth and cooler vibrations add up to about 100 kHz. 

\section{Comb preparation and two level echo efficiency}

The frequency comb required for an AFC quantum memory can be prepared using a stream of pulses where the inverse of the time separation gives the frequency separation($\Delta$) of the comb produced \cite{Usmani2010, PhysRevA.81.033803}. Spectral tailoring techniques similar to those used in references \cite{PhysRevLett.104.040503} and \cite{submittedprb} are necessary before the comb creation to isolate an atomic system such as that described in figure \ref{fig:lambda_syst}. Note that the spectral tailoring is performed over a certain frequency range, in our case over roughly 10 MHz. The spectral tailoring includes a spin polarization process, which prepares the atoms in one of the ground states, in this case in the auxiliary state $\pm|1/2\rangle_g$ (see figure \ref{fig:lambda_syst}). The comb is then prepared by sending a stream of 15 pulses on the \trans transition, exciting a spectral comb structure of ions. A fraction of the excited ions in $\pm|5/2\rangle_e$ will decay to $|3/2\rangle_g$, thus forming the desired spectral comb of ions on \tranz. Since only a fraction of the ions decay to $|3/2\rangle_g$, it is necessary to repeat the comb preparation many times. Ions that fall down in the $|5/2\rangle_g$ must also be removed by optical pumping, forcing most of these ions into the desired $|3/2\rangle_g$ state. In the experiments shown here 80 repetitions are used. More details on all the steps used in the comb preparation can be found in \cite{submittedprb}.

\begin{figure}
\centering
\includegraphics[scale=0.4]{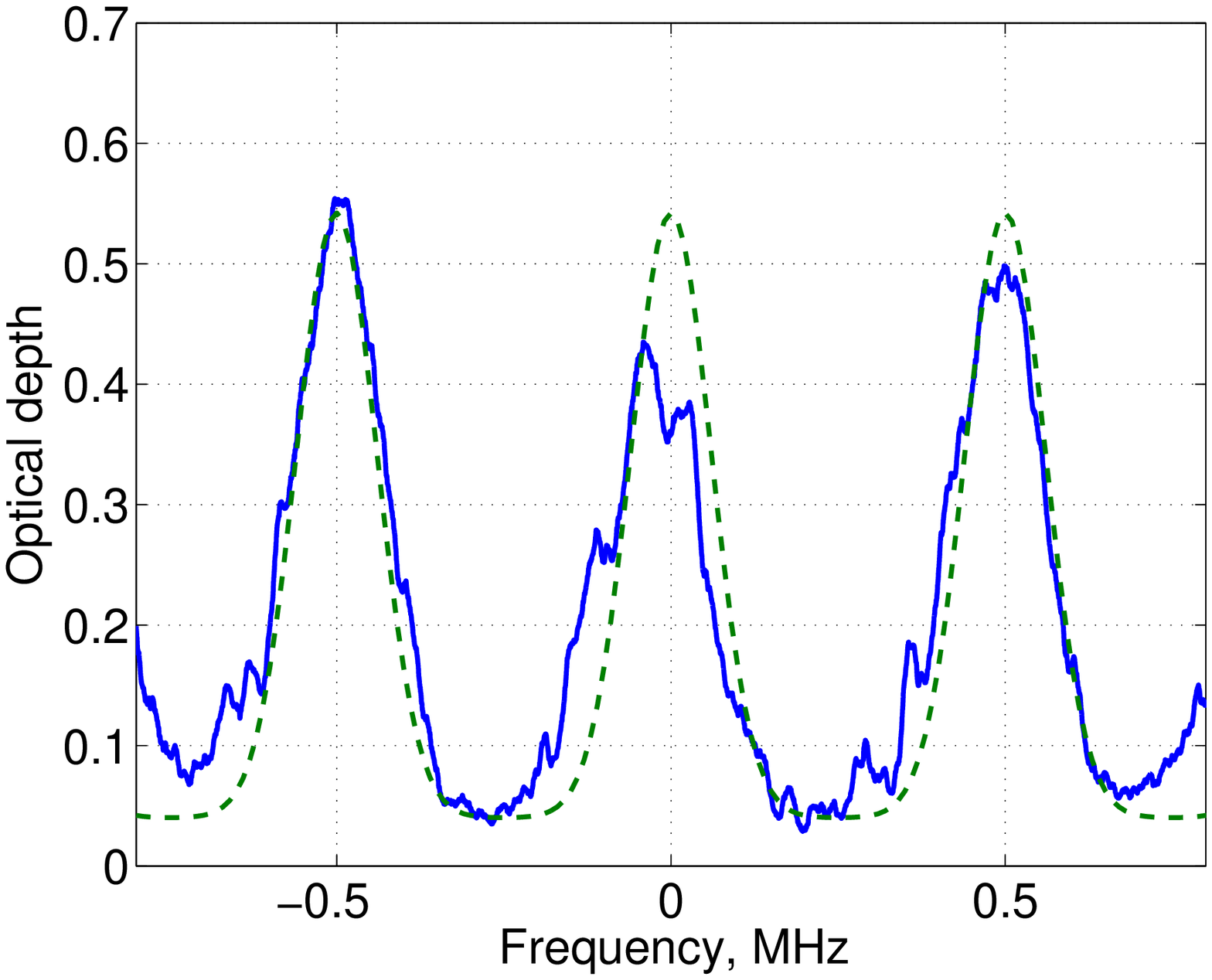}
\includegraphics[scale=0.4]{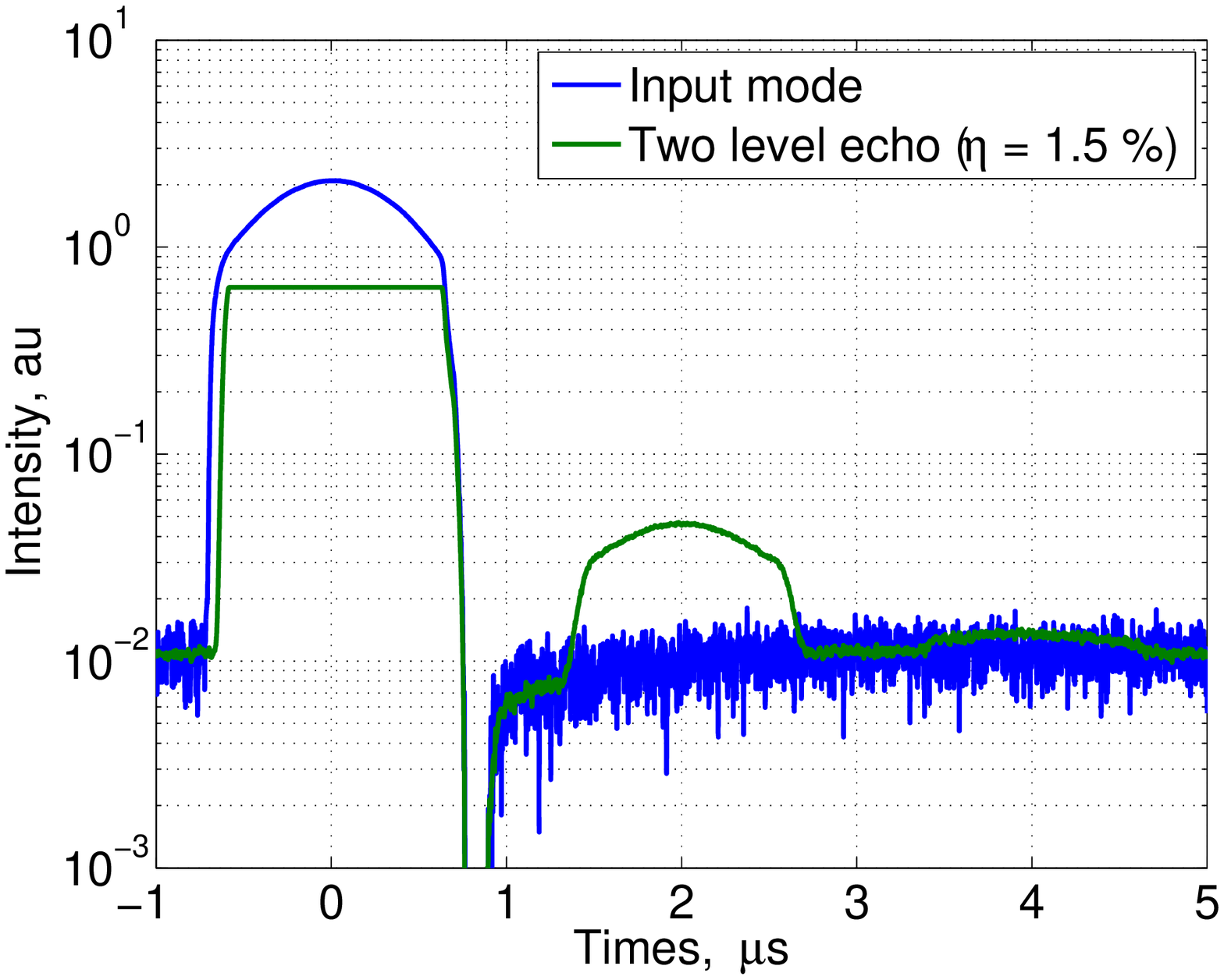}
\caption{(a) A sample comb created with a periodicity of 0.5 MHz. The blue solid line shows the measured comb. This comb is obtained by measuring the absorption of the \tranz with a small scan of $<$2 MHz over 200 $\mu$s. The comb shown is an average of 20 measurements. The green dashed line shows a Gaussian comb with $\gamma$ = 165 kHz (corresponding to F = 3.03), d$_0$ = 0.04 and d = 0.54. (b) We here show the corresponding echo, where the efficiency is measured to be 1.5 $\%$. It was necessary to change the detector gain for the two measurements shown, which results in more noise on the background of the input mode.}
\label{fig:comb2echo}
\end{figure}

A sample comb is shown in figure \ref{fig:comb2echo}a where the frequency separation of the comb is $\Delta$ = \mbox{0.5 MHz} (this corresponds to a two level AFC storage time of \mbox{2 $\mu$s}). The comb measurement is performed by measuring the absorption of \tranz. The maximum optical depth that we could obtain on this transition is $d$ = 0.8, higher than the maximum peak height shown in figure \ref{fig:comb2echo}a, where the maximum value is $d$ = 0.54. Peaks with higher absorption are obtainable with more power in the pulse stream, however, this results in broader peaks. Broader peaks reduce the finesse (F) of equation \ref{eq:GaussEff}, thus lowering the maximum echo efficiency possible. The green dotted line shown in figure \ref{fig:comb2echo}a represents a Gaussian comb in frequency ($\nu$) of the form

\begin{equation}
n(\nu) = d\sum_{j=-4}^{j = +4} e^{-\frac{(\nu-j\Delta)^2}{2{\bar{\gamma}}^2} } +d_0
\label{eq:GaussianComb}
\end{equation}

\noindent where the peak width is given by $\bar{\gamma}=\gamma/\sqrt{8 \ln 2}$ (all other parameters are defined as before). The finesse of each comb is obtained by measuring the width of the peaks created. The minimum width of a peak in the comb is limited by the laser linewidth and the frequency noise induced by vibrations from the cryostat. Additional contributions include possible power broadening from the peak preparation pulses and the inhomogeneous spin linewidth of the material.

\begin{figure}
\centering
\includegraphics[scale=0.5]{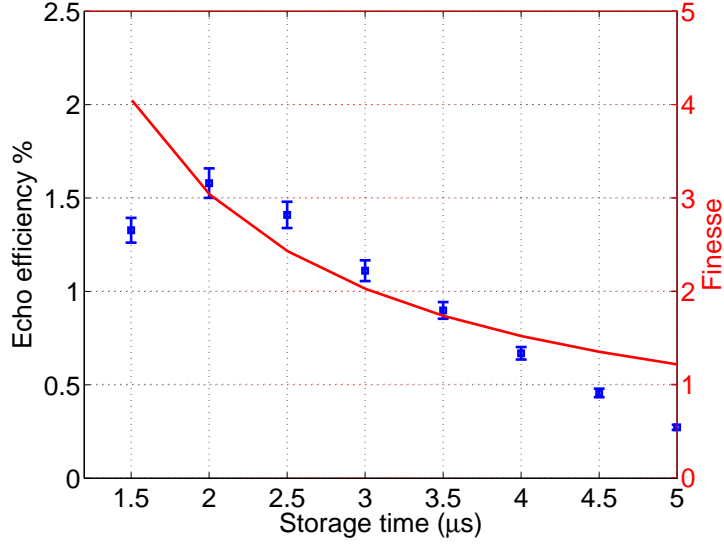}
\caption{This figure shows the decay in the echo efficiency of a two level AFC echo. The squares show the efficiencies using a comb preparation described in the text for a range of AFC storage times($1/\Delta$). The finesse expected from $\gamma$ = 165 kHz is plotted using a solid red line (right hand y axis).}
\label{fig:twoLdecay}
\end{figure}

The echo which can be seen using the comb in figure \ref{fig:comb2echo}a is shown in figure \ref{fig:comb2echo}b. The efficiency of the emitted two level echo is obtained by comparing the area of the input mode when there is no absorption on the input mode transition (the blue solid line), to the area of echo signal (the green solid line). The efficiency measured here is 1.5 $\%$, higher than that expected using equation \ref{eq:GaussEff} and the values of d and d$_0$ obtained in figure \ref{fig:comb2echo}b of 1.2 $\%$ (cf. figure \ref{fig:twoLdecay}). We estimate an error of \mbox{$\pm$5 $\%$} on the efficiencies shown due to the non linear reaction of changing the gain on the detector. Additionally, the optical depth measured in the comb measurement of figure \ref{fig:comb2echo} is likely to be too low. Firstly, the optical depth has been reduced by the input pulse which was not suppressed for this measurement. Secondly the rate of the scan, 2 MHz in 200 $\mu$s implies a scan resolution of roughly 100 kHz, large enough to detrimentally affect the values extracted for d and $d_0$.

Using the same preparation method, the time  separation of the comb preparation pulses and thus the frequency separation of the comb teeth($\Delta$) was varied. The two level echo efficiency for a storage time of up to 5 $\mu$s was measured and the results are plotted in figure \ref{fig:twoLdecay}. A comb has been measured for each efficiency shown in figure \ref{fig:twoLdecay}. For longer storage times, we do not deem this comb measurement to be accurate. The effects of our scan resolution, the vibrating cryostat and the laser linewidth are hard to separate from the actual comb. Instead we merely extract a value of $\gamma$ from these plots of 165 kHz, and show the measured finesse on the right hand axis. The trend of decreasing finesse accounts for the trend of decreasing efficiency.

The optimum storage time in our system is currently 2 $\mu$s. But these results do not represent a fundamental limit to the maximum storage time or two level efficiency in this material. The storage time is clearly limited by the AFC tooth width ($\gamma$). The minimum width obtained was limited by our laser linewidth and the effective linewidth broadening probably due to crystal movement induced by the cryostat cooler. In order to improve the storage time we should improve the laser frequency stabilization and reduce the presumed effect of crystal movement. The latter can be done by employing a low-vibration pulse tube cooler or by stabilizing the path length between the laser source and crystal, this could be done using an interferometric setup. In addition, due to our preparation method, the inhomogeneous spin linewidth is also likely to contribute towards the minimum linewidth. This issue can, however, easily be resolved by changing the preparation method.

The maximum efficiency obtained at 2 $\mu$s storage time is limited by the low optical depth of our material. Higher efficiencies can certainly be reached using a crystal with higher doping concentration. Indeed a peak optical depth three times higher has been obtained \cite{PhysRevB.68.085109}, which ideally would result in a ten-fold increase in the efficiency (see eq. \ref{eq:GaussEff}). Another interesting solution would be to place a cavity around the crystal \cite{PhysRevA.82.022310, PhysRevA.82.022311}. Both methods might be required in order to approach unit efficiency. We conclude this section by noting that whilst not dictating an absolute limit, these efficiencies do represent a serious limitation in our present system, which is relevant to the following results of spin wave storage in the complete AFC scheme.

\section{Spin wave storage}

\begin{figure}
\centering
\includegraphics[scale=0.5]{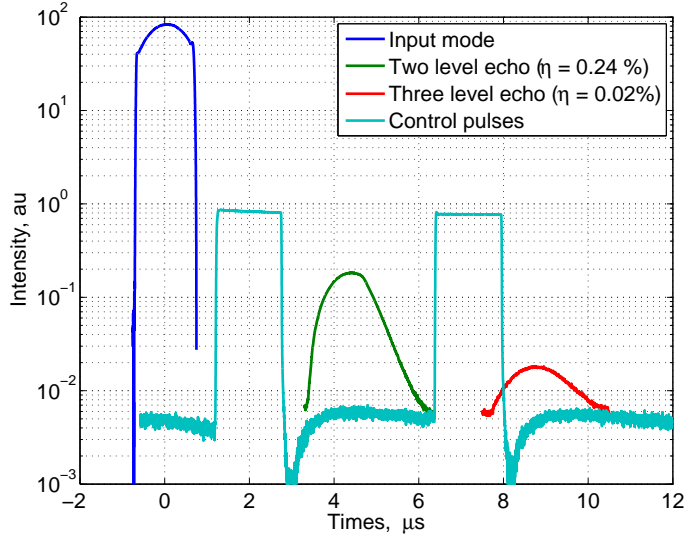}
\caption{This figure shows an input mode (blue), its two level echo with no control pulses (green) and a spin stored echo, a three level echo when the control pulses are applied (red). Compared to the two level echoes shown in figure \ref{fig:twoLdecay} the input mode is not in the same spatial mode as the preparation of the AFC. This is to reduce free induction decay background noise on the three level echo. The input mode and the preparation beams overlap in the crystal. The control fields are in the same mode as that of the preparation. Imperfect overlap of the two modes accounts for the efficiency of the two level echo dropping by almost a factor of three compared to figure \ref{fig:twoLdecay}. Changing the gain on the detector distorts the pulse shape of the echo. All of the measurements represent an average of 20 traces.}
\label{fig:level3echo}
\end{figure}

In order to perform spin wave storage, we spatially separate the input mode from the control mode using a cross beam configuration, as already discussed in the experimental description section(see also figure \ref{fig:schem}). While reducing the free induction decay noise caused by the control fields \cite{PhysRevLett.104.040503}, the crossed beam configuration also has a negative effect, it reduces the efficiency of the two level echo. A sample two level echo is shown in figure \ref{fig:level3echo}. This echo is measured without the control pulses, but in a crossed mode configuration where the comb is prepared using AOM:M. The efficiency of the two level echo is measured to be \mbox{0.24 $\%$} at \mbox{$1/\Delta = 4\mu$s}. This should be compared to the single spatial mode measurement, where the efficiency was measured to be \mbox{0.6 $\%$} (see figure \ref{fig:twoLdecay}). The two level echo efficiency is thus reduced by almost a factor of 3. This is attributed to imperfect overlap of the two beams.

In the introduction of this paper we talked about using chirped pulses for our control pulses. Such pulses would allow us to efficiently transfer a large bandwidth, potentially all of the teeth in the AFC spectrum. But these pulses are generally of longer duration as compared to a $\pi$-pulse \cite{PhysRevA.82.042309}. There is however a limit on the time which we can use to perform the control pulses. The duration and shape of the input pulse and one control pulse cannot be longer than the predefined two level echo time $1/\Delta$. In addition, the duration of the input pulse defines the bandwidth which the control pulse must transfer. In our current experimental set up, the Rabi frequency of the control pulse transition, \trann, is estimated to be of the order of 300 2$\pi$kHz. To transfer a bandwidth of 300 kHz, using a $\pi$-pulse, 1.7 $\mu$s are necessary at this Rabi frequency. If we set $1/\Delta$ = 2 $\mu$s to obtain the highest two-level echo efficiency, the input pulse would have to be shorter than 0.3 $\mu$s. The spectral bandwidth of such an input pulse would be $>$3 MHz, the majority of which will not be transferred by a 1.7 $\mu$s control pulse. This illustrates that a compromise must be made between the efficiency of the two level echo and the efficiency of the transfer. We use the final echo efficiency as an indicator of the best compromise.

The input pulse had approximately a Gaussian shape, while the identical control pulses were square shaped. We set the two-level storage time to $1/\Delta=4\mu$s. The durations of the input pulse and the control pulses were optimized by looking at the efficiency of the three level echo. The highest efficiency was reached with an input pulse with a full width at half maximum of \mbox{1.3 $\mu$s}, and control pulses of duration 1.55 $\mu$s. The resulting echo with spin-wave storage is shown in figure \ref{fig:level3echo}, where the spin storage time is roughly 5 $\mu$s, leading to a total storage time of 9 $\mu$s. We checked that the echo is not present if we remove the first control pulse or if we remove the input mode, as expected. In figure \ref{fig:level3echo} we also show the associated two-level echo (no control pulses applied) obtained with an identical comb structure. Note that AOM:D is used to gate the detector around the measured echoes, as the input mode would saturate the detector at the gain required to see the echoes.

\begin{figure}
\centering
\includegraphics[scale=0.5]{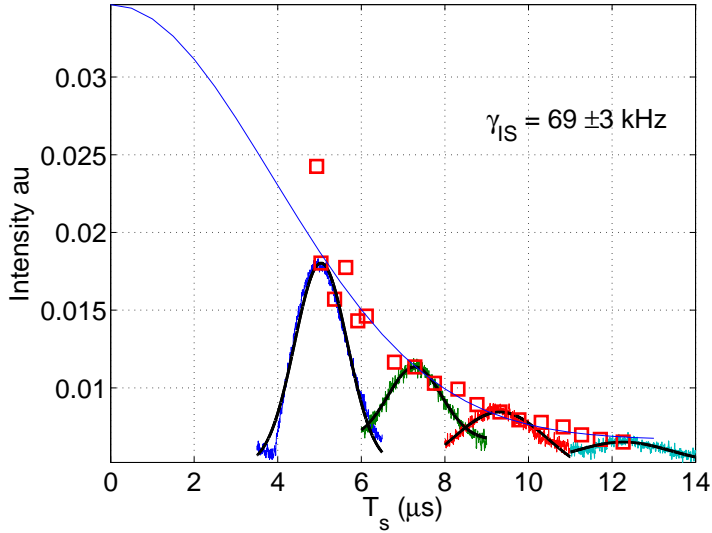}
\caption{Inhomogeneous spin linewidth measurement. Each point represents the maximum of the Gaussian function which has been fitted to each trace. Some traces and fits are shown, the rest are suppressed for clarity. The formula used to fit the data is described in the text, yields an inhomogeneous linewidth of \mbox{69 $\pm$ 3 kHz}.}
\label{fig:inhomdec}
\end{figure}

Maintaining this \mbox{$1/\Delta=4\mu$s} AFC  it is possible to vary the time between the control pulses ($T_s$). This is the on demand storage time of the full AFC scheme. For a rare earth doped crystal we expect to see a decay given by the inhomogeneous spin linewidth of the material \cite{PhysRevLett.104.040503}. The storage time of the complete AFC scheme as described in the introduction of this paper is limited by this linewidth. If we assume a Gaussian distribution for the spin broadening the decay of the stored echo is given by

	\begin{equation}
		\mbox{Echo height} = A e^{\left(\frac{-T_s^2\gamma_{IS}^2\pi^2}{2ln2}\right)}
		\label{eq:decay3}
	\end{equation}

\noindent where $T_s$ defines the spin storage time, $A$ is a constant and $\gamma_{IS}$ is the inhomogeneous spin linewidth. An echo of increasing spin storage time is shown in figure \ref{fig:inhomdec}. The complete memory time is given by $T_s+1/\Delta$. Each recorded echo is fitted with a Gaussian function, the maximum of which is used to fit equation \ref{eq:decay3}. We obtain an inhomogeneous spin linewidth of \mbox{$69 \pm 3$ kHz}. Some echo traces and their fits are shown, but most have been suppressed for the sake of clarity.

Armed with the knowledge of the inhomogeneous spin linewidth we can estimate the efficiency of the transfer pulses used. For the example shown in figure \ref{fig:level3echo}, the measured efficiency can be extrapolated to $T_s = 0$, resulting in an estimated 0.04$\%$ efficiency without spin dephasing. Hence, the estimated three level echo at $T_s = 0$ is 16 $\%$ of the two level echo (0.04\%/0.24\%=0.16). Following the simple model discussed in \cite{PhysRevLett.104.040503} we can then calculate the transfer efficiency per pulse, which we find to be 40\%. It should be noted that since the bandwidth of the control pulses is not much larger than the input pulse bandwidth, we should take this value as an effective average over the bandwidth. It is also difficult to estimate the effect of imperfect beam overlap on the efficiency. However, considering we observe a strong decrease in the two level echo due to insufficient overlap, it is likely that the overlap has a non-negligible effect on the estimated transfer pulse efficiency.

We emphasize that it is possible to refocus the spin coherence using radio frequency pulses (spin echo), which would allow us to store for durations of the order of the spin coherence time. This was measured to be 15.5 ms for the $^{151}$Eu$^{3+}$ isotope in Y$_2$SiO$_5$ at zero magnetic field \cite{Alexander2007}. One can expect a similar value for $^{153}$Eu$^{3+}$. As a future perspective one can also think of applying a magnetic field to cancel the first order Zeeman effect, creating a memory which for europium could be of the order of many seconds \cite{PhysRevB.74.195101}. This is similar to work performed in praseodymium \cite{PhysRevLett.95.063601}. The excellent spin coherence properties is a clear strength of europium doped materials.

\section{Conclusion, discussion and outlook}

Here we show for the first time an AFC memory with spin wave storage in a europium crystal. Europium is an interesting material for quantum memories thanks to its long spin coherence time and its potential for multimode storage. The experiments we report here represent a first step in this direction. Yet, several important aspects of the memory must be significantly improved.

First, the efficiencies which we report are very low, indeed europiums' handicap lies in its low optical depth. There exist promising proposals to increase the optical depth by placing the crystal in a cavity thus increasing the efficiency of a two level echo. The measurements shown in this paper do not represent a fundamental limit of the efficiency nor of the storage time of a two level echo. The latter has an impact on the multimode capacity of the memory. The storage time could be improved by changing the comb preparation method to one such as that found in reference \cite{Usmani2010}, which would remove the effect of the inhomogeneous linewidth from the minimum peak width obtainable. Indeed peaks as narrow as 1 kHz can be found in the recent publication \cite{Thorpe2011}. Further technical improvements on the laser linewidth or a cryostat with less vibration would also improve the storage time shown in this paper, allowing for multimode storage and the use of more efficient, chirped control pulses.

Another technical difficulty: the relatively small dipole moment of this transition, means that large amounts of power are required to transfer a relatively small bandwidth. To increase the bandwidth of the transfer significantly larger amounts of power are required. Recent developments of powerful lasers for sodium based systems at 589 nm \cite{Taylor:09}, can also be applied to developing narrowband, powerful and compact lasers at 580 nm.

In the spin wave storage experiments presented here we could store for up to about 10 $\mu$s, limited by inhomogeneous spin dephasing. From these measurements we could estimate the spin linewidth, which we found to be \mbox{$69 \pm 3$ kHz}. The inhomogeneous spin linewidth is not a restriction, however, on the maximum storage time of the medium. The storage time can be increased using spin refocussing pulses such as \cite{PhysRevLett.95.063601}.

An important milestone for quantum memories based on rare-earth doped crystals would be to store an optical pulse on the single photon level as a spin wave excitation, a milestone that has not yet been reached in any rare earth based memory. The low overall efficiencies obtained in these experiments currently make this a very challenging experiment in ~\eu. The possible improvements that we have detailed, however, should make it possible in the near future. In this context we recently proposed a method of generating a photon pair source with variable delay \cite{PhysRevA.83.053840}, using the same resources used for spin-wave storage based AFC quantum memory. The overall efficiency of this source also has a less strong dependence on optical depth and control pulse transfer efficiency. An advantage of ~\eu~ in this context is the large hyperfine splitting, which facilitate spectral filtering as compared to praseodymium doped Y$_2$SiO$_5$.

\section{Acknowledgements}
We would like to thank Claudio Barreiro for technical assistance and N. Sangouard and H. de Riedmatten for useful discussions. We are also grateful to Y. Sun, R. L. Cone and R. M. Macfarlane for kindly lending us the $^{153}$Eu$^{3+}$
doped Y$_2$SiO$_5$ crystal. This work was financially supported by the Swiss NCCR-QSIT and by the European projects QuRep, Q-Essence and CIPRIS (FP7 Marie Curie Actions).

\section{Bibliography}

\bibliographystyle{unsrt}	
\bibliography{myrefs}

\end{document}